\documentclass{PoS}
\usepackage{graphicx}

\newcommand{\nh}{\ensuremath{\mbox{cm}^{-2}}}

\newcommand{\kev}{\ensuremath{\,\mbox{\scriptsize keV}}}

\newcommand{\flux}{\ensuremath{\mbox{ergs~cm}^{-2}\mbox{~s}^{-1}}}

\newcommand{\hr}{\ensuremath{HR}}
\newcommand{\nhsym}{\ensuremath{N_{\mbox{\scriptsize H}}}}
\newcommand{\sun}{\ensuremath{\odot}}

\newcommand{\swift}{{\emph{Swift}}}
\newcommand{\pedix}[2]{\ensuremath{#1_{\,\mbox{\scriptsize #2}}}}

\newcommand{\errUD}[2]{\ensuremath{^{+#1}_{-#2}}}
\newcommand{\mrk}{Mrk~915}

\def\apj{ApJ }
\def\apjl{ApJL }
\def\apjs{ApJS }

\def\aap{A\&A }

\def\mnras{MNRAS }

\title{Mining the XRT archive to probe the X-ray absorber structure in the AGN population}

\ShortTitle{AGN absorber structure through the XRT archive}

\author{\speaker{Lucia Ballo}\\
Osservatorio Astronomico di Brera (INAF), via Brera 28, I-20121, Milano (Italy); E-mail: \email{lucia.ballo@brera.inaf.it}}
\author{Paola Severgnini, Alberto Moretti, Roberto Della Ceca, Stefano Andreon, Alessandro Caccianiga\\
Osservatorio Astronomico di Brera (INAF), via Brera 28, I-20121, Milano (Italy)}
\author{Valentina Braito\\
Osservatorio Astronomico di Brera (INAF), via E. Bianchi 46, I-23807 Merate, LC (Italy);\\
Department of Physics, University of Maryland, Baltimore County, Baltimore, MD 21250 (USA)}
\author{Sergio Campana\\
Osservatorio Astronomico di Brera (INAF), via E. Bianchi 46, I-23807 Merate, LC (Italy)}
\author{Cristian Vignali\\
Dipartimento di Fisica e Astronomia, Universit\`a degli Studi di Bologna, viale Berti Pichat 6/2, I-40127, Bologna (Italy);\\
Osservatorio Astronomico di Bologna (INAF), via Ranzani 1, I-40127, Bologna (Italy)}

\abstract{
One of the key ingredients of the Unified Model of Active Galactic Nuclei (AGN) is the presence of a torus-like optically thick medium composed by dust and gas around the 
putative supermassive black hole. 
However, the structure, size and composition of this circumnuclear medium are still matter of debate. 
To this end, the search for column density variations through X-ray monitoring on different timescales (months, weeks and few days) is fundamental to constrain size, kinematics and location of the X-ray absorber(s).

Here we describe our project of mining the  \swift-XRT archive to assemble a sample of AGN with 
extreme column density variability 
and determining the physical properties of the X-ray absorber(s).
We also present the results obtained from a daily-weekly \swift-XRT follow-up monitoring recently performed on one of the most interesting new candidates for variability discovered
so far, \mrk.
}

\FullConference{Swift: 10 Years of Discovery,\\
		2-5 December 2014\\
		La Sapienza University, Rome, Italy }

\begin{document}

\section{Variable \swift-AGN sample}
\vspace{-0.3cm}
According to the Unified Model of Active Galactic Nuclei (AGN; \cite{antonucci93}), an obscuring optically thick medium composed by dust and gas arranged in a torus-like geometry 
is present around the nuclear engine. 
However, the structure, size and composition of this circumnuclear medium are still matter of debate; 
important constrains have been recently provided by the study of the absorption variability, 
almost ubiquitous in bright absorbed AGN (\cite{risaliti02}).
X-ray absorbing column densities, \nhsym, have been observed to change over different timescales, from a few years down to weeks or days, for a handful of AGN 
(e.g., \cite{mark14} and references therein),
allowing us to investigate the X-ray properties of the absorbers down to sub-parsec scales.
The emerging picture is that multiple neutral and ionized absorbers co-exist around the central black hole, located at different distances from it.

We recently started a project aimed at finding more examples of these sources in order to study the physical properties of the X-ray absorbers on a larger statistical basis.
We exploited the vaste amount of X-ray observations provided by the X-ray Telescope (XRT; \cite{xrt}) onboard the satellite \swift.
XRT long exposure observations are always split in segments of $\sim 1\,$ks, with typical integration times of $5\,$ks per day. 
This generates, for bright sources, a natural short-timescale monitoring, with spectra good enough to perform a basic spectral analysis.
As an example, for an AGN with a flux of $\pedix{F}{0.5-10\kev}\sim 5\times 10^{-12}\,$\flux, in $\sim 5\,$ks we expect to collect $\sim 500$\,net photons.
This makes the \swift-XRT archive an important database to search for different timescale variability of bright sources.
To assemble our sample, we considered: 
\begin{enumerate}
\vspace{-0.2cm}
 \item the serendipitous sources detected in the field of view of XRT observations performed as follow-up of 76 gamma-ray bursts.
 These observations correspond to $\sim49$~square degrees;
\vspace{-0.2cm}
 \item the XRT sources classified as AGN that are possible counterparts of objects detected by the Burst Alert Telescope (BAT) onboard \swift\ \cite{batcat,cusumano10}.
This AGN sample can be considered unbiased in term of column density, at least for $\nhsym\lesssim 10^{24}\,$\nh\ (a limit above which a source falls in the so-called Compton-thick regime).
In the energy band covered by BAT ($14-195\,$keV), the intrinsic emission is indeed weakly affected by absorption levels below few $10^{24}\,$\nh.
\end{enumerate}
\vspace{-0.2cm}

\noindent 
For each source, 
we split the XRT data sets on a daily basis, combining all the observation segments within 1 day (for the AGN-BAT sample) or 2 days
(for the serendipitous sources).
We filtered out time bins shorter than $200\,$sec.
The light curves span a wide range of number of time bins (from one up to more than $200$ spectra) and timescales (from few days to years).
For $299$ serendipitous sources and $405$ BAT AGN we have at least two spectra.

\begin{figure}
 \begin{minipage}[b]{0.455\linewidth}
 \centering
   \includegraphics[width=1\textwidth]{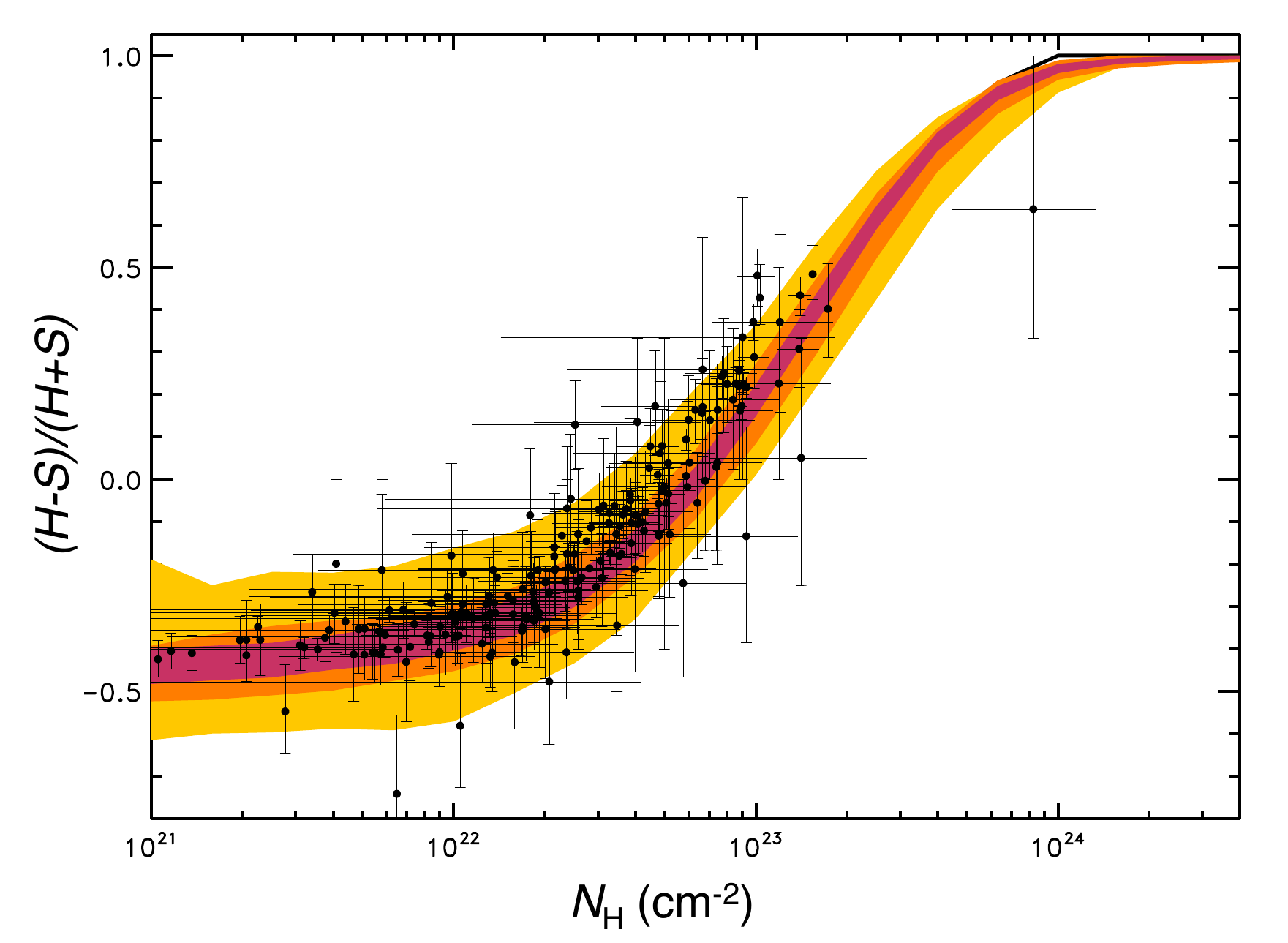}
   \end{minipage} 
 \begin{minipage}[b]{0.545\linewidth}
 \centering
    \includegraphics[width=1\textwidth]{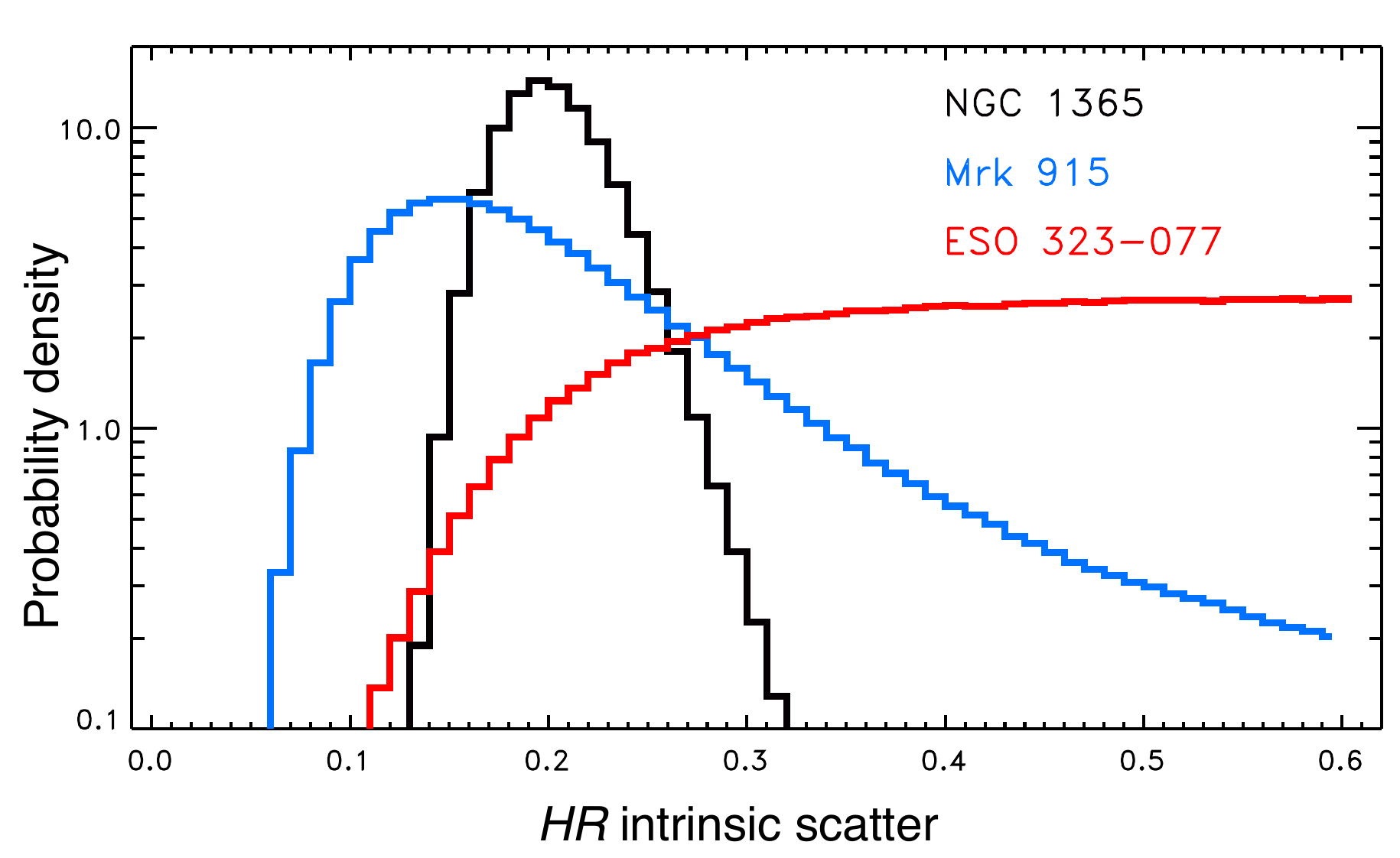}
  \end{minipage} 
\vspace{-0.9cm}
\caption{\footnotesize
{\it Left panel:} \hr\ of simulated spectra ($\Gamma=2$) for different values of $N_{\mbox{\tiny H}}$. 
Shaded bands: $16^{\mbox{\tiny th}}$ and $84^{\mbox{\tiny th}}$ percentile from 100 simulations assuming source counts from 20 (yellow) to 500 (purple). 
Superimposed, we plot the real values derived by the XRT data for the BAT AGN considered here (black dots and error bars).
{\it Right panel:} Examples of posterior probability of \hr\ intrinsic scatter for 3 sources for which spectral variability has been observed at 
$99.99...$\% confidence level.
}
\vspace{-0.3cm}
\label{fig:prob}
\end{figure}

\begin{figure}
 \begin{minipage}[b]{0.5\linewidth}
 \centering
    \includegraphics[width=0.9\textwidth]{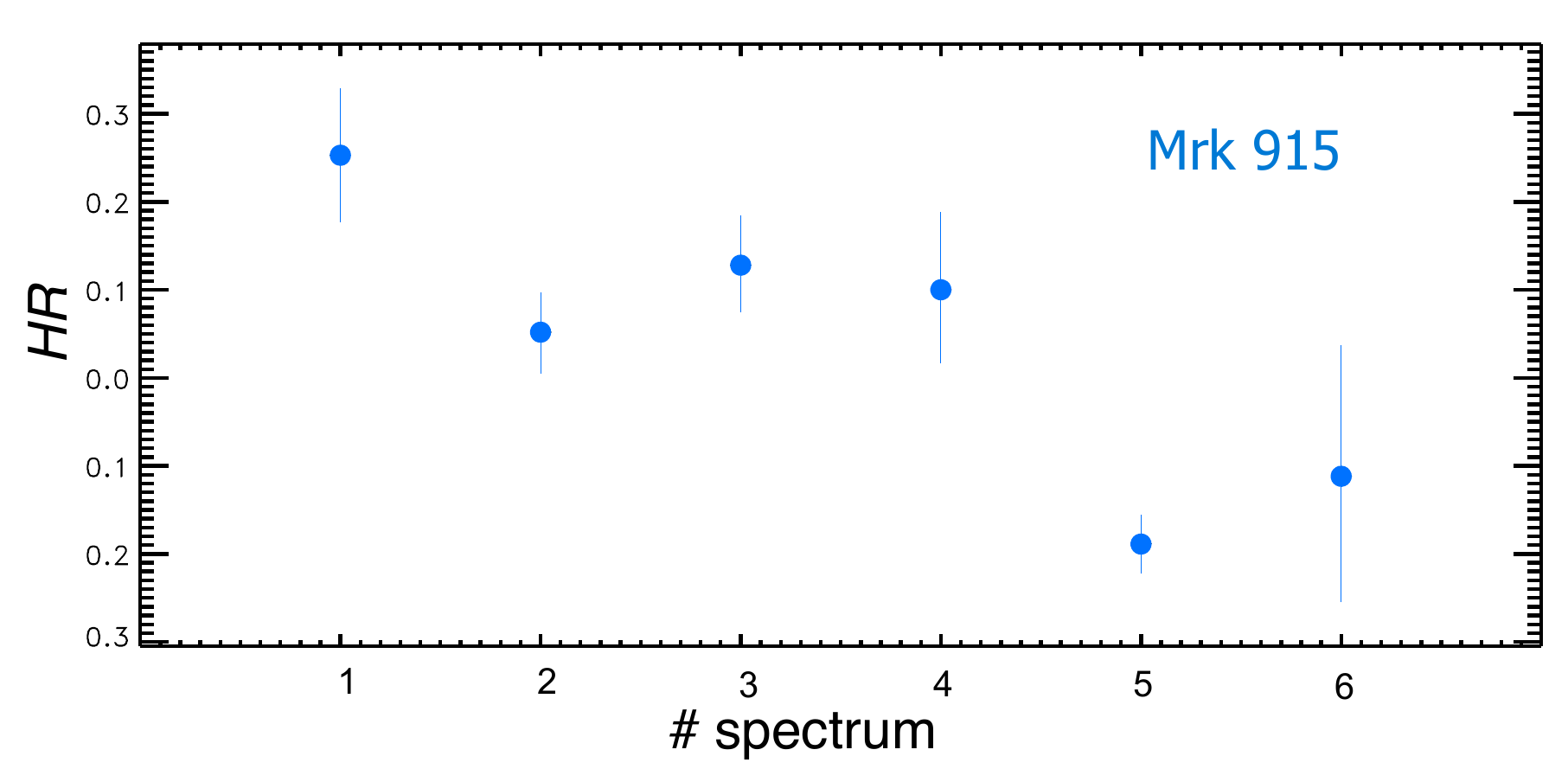}
   \end{minipage} 
 \begin{minipage}[b]{0.5\linewidth}
 \centering
    \includegraphics[width=0.9\textwidth]{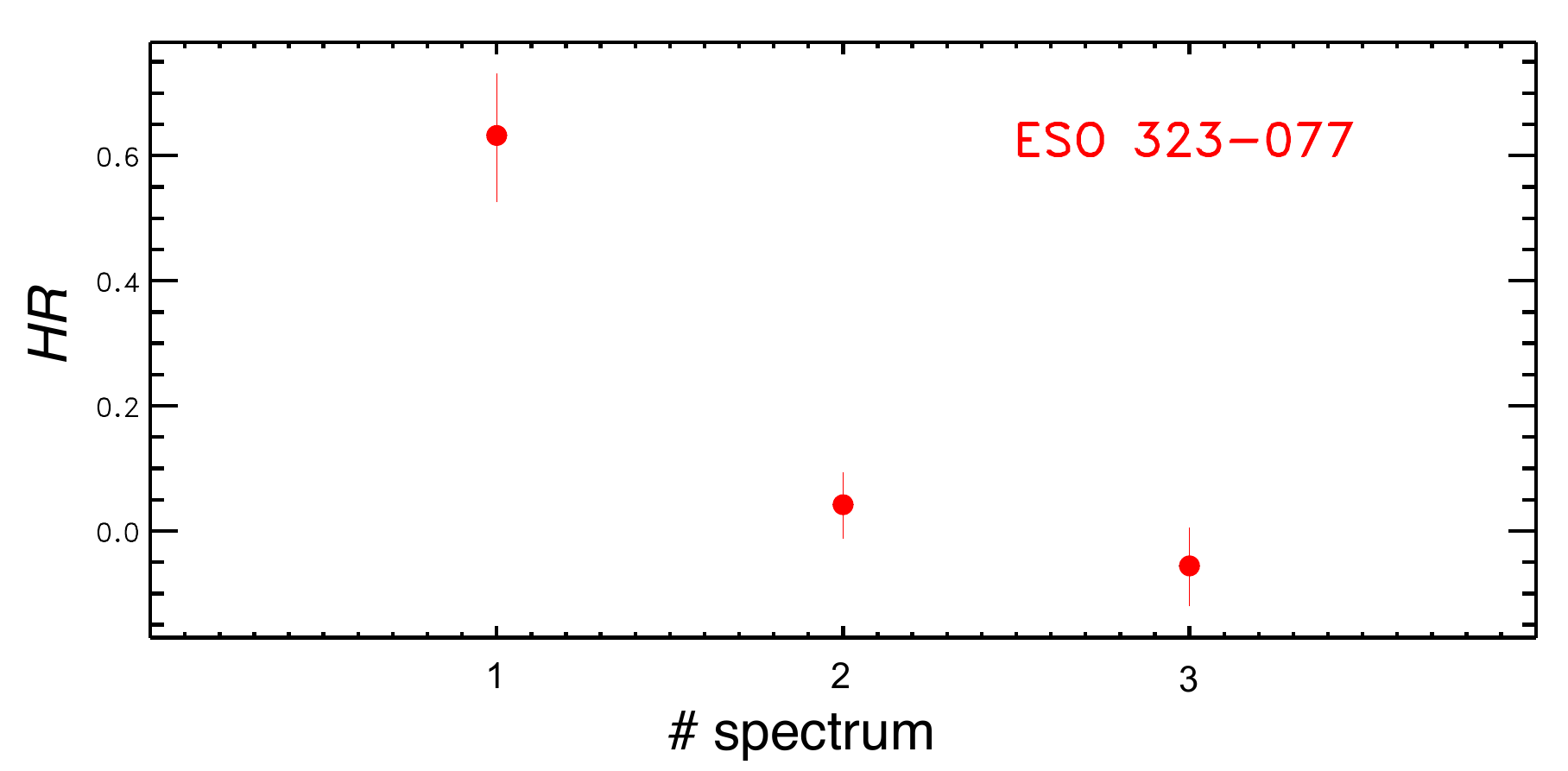}
  \end{minipage} 
\vspace{-0.8cm}
\caption{\footnotesize
Observed \hr\ light curves derived from the XRT archival data binned on a daily basis for two of the variable candidates (\mrk, {\it left};  and ESO~323-077, {\it right}). 
}
\vspace{-0.5cm}
\label{fig:lc}
\end{figure}

As a proxy of the amount of column density we used the hardness ratio $\hr=(H-S)/(H+S)$, where $S=2-4\,$keV and $H=4-10\,$keV.
In the $2-10\,$keV energy band, the spectrum of an AGN can be approximate, at the first order, by an absorbed power law, while
below $\sim 2\,$keV different soft components (soft excess, reflection, scattering, etc) are also present.
Under this hypothesis, the \hr\ value as defined here provides a strong indication of the amount of absorption:
in Fig.~\ref{fig:prob} (left panel) we compare the \hr\ vs. \nhsym\ derived from the analysis of the XRT data with the 
\hr\ measured for a number of simulated spectra (photon index $\Gamma=2$) with different values of absorbing column density.
For each input \nhsym\ value, we indicate with different colours the $16^{\mbox{\footnotesize th}}$ and $84^{\mbox{\footnotesize th}}$ percentile of the 
distribution of $100$ tests with three different values of $2-10\,$keV source photons 
(from lightest to darkest, $20$, $100$ and $500$).
From the plot it is evident that the \hr\ 
is a proxy of the \nhsym, and variation in the \hr\ can be interpreted as an indication of a change in the absorbing column density, 
at least for \nhsym\ between $\sim10^{22}\,$\nh\ and few times $10^{23}\,$\nh.

In order to probe the spectral variability, 
for each source we measured the possible non-zero amplitude of an intrinsic \hr\ scatter 
superimposed to the statistical one. 
We adopted a Bayesian approach similar to the one used by \cite{andreon08} to determine the intrinsic scatter of cluster velocity dispersion in
presence of measurement errors, adapted here to Poisson-based quantities.
Basically, starting from the source and background photons in the two bands ($2-4\,$keV and $4-10\,$keV) and in the different time bins, 
we calculate
the probability for each value of intrinsic scatter given the data.
As an example, the  probability distributions 
for $3$ sources for which a spectral variation is clearly detected are shown in Fig.~\ref{fig:prob} (right panel), where
we plot the probability associated to an intrinsic scatter value ranging from $0$ (corresponding to the no-scatter case) to $0.6$.
As evident from the figure, the probability of null intrinsic scatter is $\ll 10^{-3}$, implying a negligible probability that the measured \hr\ variability 
(shown in Fig.~\ref{fig:lc} for two of the sources) is due to statistical fluctuations.

By applying a very conservative threshold of $10^{-3}$ in the probability of null intrinsic scatter, we found $15$ AGN candidates for variable absorbers, 
$2$ of which are serendipitous and $13$ are BAT-selected.
The sources are listed in Table~\ref{tab:varcand}, together with the main characteristics of the XRT archival data considered here.
For the AGN observed by BAT, we summarise also the basic information extracted from the \swift-BAT 70 month survey catalogue\footnote{See http://swift.gsfc.nasa.gov/results/bs70mon/}.
Most of the BAT-selected sources are well-known variable AGN: this confirms the validity of the adopted method.
We are exploring its potentiality by relaxing the threshold in probability of null intrinsic scatter.
However, even the more restrictive threshold used here demonstrates the effectiveness of this method: it provided us with 
$5$ new candidates (two of which are serendipitous) that were not previously known as variable sources.
Below we present the results obtained from a daily-weekly \swift-XRT follow-up monitoring that was recently performed on one of 
them, \mrk.

\begin{table}
\begin{center}
{\scriptsize
\begin{tabular}{l |c@{\extracolsep{0.2cm}}  c@{\extracolsep{0.2cm}}  c || c@{\extracolsep{0.2cm}}  c@{\extracolsep{0.2cm}}  c@{\extracolsep{0.2cm}} c || @{\extracolsep{0.2cm}}c@{\extracolsep{0.2cm}}}
 \multicolumn{1}{c|}{Source name} & \multicolumn{3}{c||}{XRT archive} & \multicolumn{4}{c||}{BAT catalogue} & \multicolumn{1}{c}{Ref.} \\
  \cline{2-8}
  & Tot. Exp. & $\langle$count rate$\rangle$ & \# spectra  & $F_{14-195\,\mbox{keV}}$ & $\Gamma_{14-195\,\mbox{keV}}$ & {\it z} & Opt. class. & \\
  \multicolumn{1}{c|}{(1)} & (2) & (3) & (4) & (5) & (6) & (7) & (8) & (9) \\
\hline \hline
 \mrk              &  23.2 &  0.14 &    6       & 31$\pm$6 & 1.7$\pm$0.3 & 0.02411 & Sy1 & \\
 NGC 1365             & 154.8 &  0.14 &   34 & 64$\pm$4 & 2.0$\pm$0.1 & 0.00546 & Sy1.8 &  {\tiny \cite{risaliti05,risaliti07,maiolino10,braito14,torricelli14}} \\
 ESO 362-18           &  96.9 &  0.53 &   31 & 49$\pm$4 & 1.9$\pm$0.1 & 0.01245 & Sy1.5 & {\tiny \cite{agis14}} \\
 CGCG 031-072         &  10.7 &  0.03 &  3 & 21$\pm$5 & 1.9$\pm$0.3 & 0.03305 & Sy1 & \\
 NGC 4395             & 243.6 &  0.11 & 190 & 25\errUD{4}{3} & 2.0$\pm$0.2 & 0.00106 & Sy1.9 & {\tiny \cite{gofford13,torricelli14}} \\
 ESO 323-077          &   7.5 &  0.09 &     3 & 33 & 2.1 & 0.01501 & Sy1.2 & {\tiny \cite{miniutti14}} \\
 Mrk 335              & 299.2 &  0.34 &    182 & 18$\pm$4  & 2.3$\pm$0.3 & 0.02578 & Sy1.2 & {\tiny \cite{grupe08,grupe12,longinotti13,mark14}} \\
 NGC 3227             &  14.5 &  0.32 &      6 &  10$\pm$4 & 2.0$\pm$0.1 & 0.00386 & Sy1.5 & {\tiny \cite{gofford13,mark14,torricelli14}} \\
 NGC 3516             &  36.8 &  0.46 &  31 & 118$\pm$4 & 1.9$\pm$0.1 & 0.00884 & Sy1.5 & {\tiny \cite{turner11,huerta14,mark14,torricelli14}} \\
 NGC 4051             &  39.6 &  0.67 &  32 & 40$\pm$3 & 2.3$\pm$0.1 & 0.00233 & Sy1.5 & {\tiny \cite{uttley04,lobban11,gofford13,mark14,torricelli14}} \\
 Mrk 766              & 108.5 &  0.58 &    56 & 22$\pm$3 & 2.5$\pm$0.2$\pm$ & 0.01293 & Sy1.5 & {\tiny \cite{pounds03,miller07,turner07,risaliti11,gofford13}} \\
 NGC 3783             &  21.7 &  2.2 &     6  & 181$\pm$5 & 2.0$\pm$0.1 & 0.00973 & Sy1 & {\tiny \cite{reis12,gofford13,mark14,torricelli14}} \\
 MCG +11-11-032  &  14.6 &  0.02 &    3 & 18$\pm$4 & 1.8$\pm$0.3 & 0.036 & Sy2 & \\
\hline
   &  &  &   \\
 SWJ012118-125727 &  53.6 &  0.01 &    4 \\
 SWJ105826-231414 &  59.4 &  0.004 &    7 \\
\end{tabular}
 }
 \end{center}       
 \vspace{-0.5cm}
\caption{\footnotesize   \swift-based information for the $15$ objects that our method candidates as variable with high confidence.
We list in the first part the $13$ BAT-selected AGN, and in the second part the $2$ serendipitous sources.
 Col. (1): Source name.
 Cols. (2) - (4): Information related with the XRT archival data used in this work: total exposure time, in units of ksec; average $0.5-10\,$keV count rate; 
 number of useful spectra.
 Cols. (5) - (8): Information derived from the \swift-BAT 70 month survey catalogue: 
 BAT flux in the $14-195\,$keV energy band, in units of $10^{-12}\,$ergs~cm$^{-2}$~s$^{-1}$; spectral index (computed from a power-law fit to the 8-band BAT data); 
 redshifts (taken from the online databases NED and SIMBAD); optical classification.
 Col. (9): Not-exhaustive list of previous studies of variability in the X-ray band.}
\vspace{-0.5cm}
\label{tab:varcand}
\end{table}%

\vspace{-0.3cm}
\section{A new candidate for variable absorber: \mrk}
\vspace{-0.3cm}
\mrk\ ($z=0.024$) is a Seyfert 1.5-1.9 galaxy hosting a black hole of mass $\pedix{M}{BH}=0.6-1.8\times10^8\,\pedix{M}{\sun}$ (\cite{bennert06}).
Spectroscopic observations performed in the optical band (\cite{bennert06,trippe10}) suggest that part of the optical absorption affecting the central regions originates outside the obscuring torus, 
and it is most likely associated with the dust lanes seen crossing the central source (\cite{malkan98}).
For this source the \hr\ variation observed 
has a probability less than $10^{-6}$ of having a null intrinsic scatter superimposed to the statistical fluctuations 
(Fig.~\ref{fig:lc}, left panel). 

\begin{figure}
\centering
    \includegraphics[width=1.\textwidth]{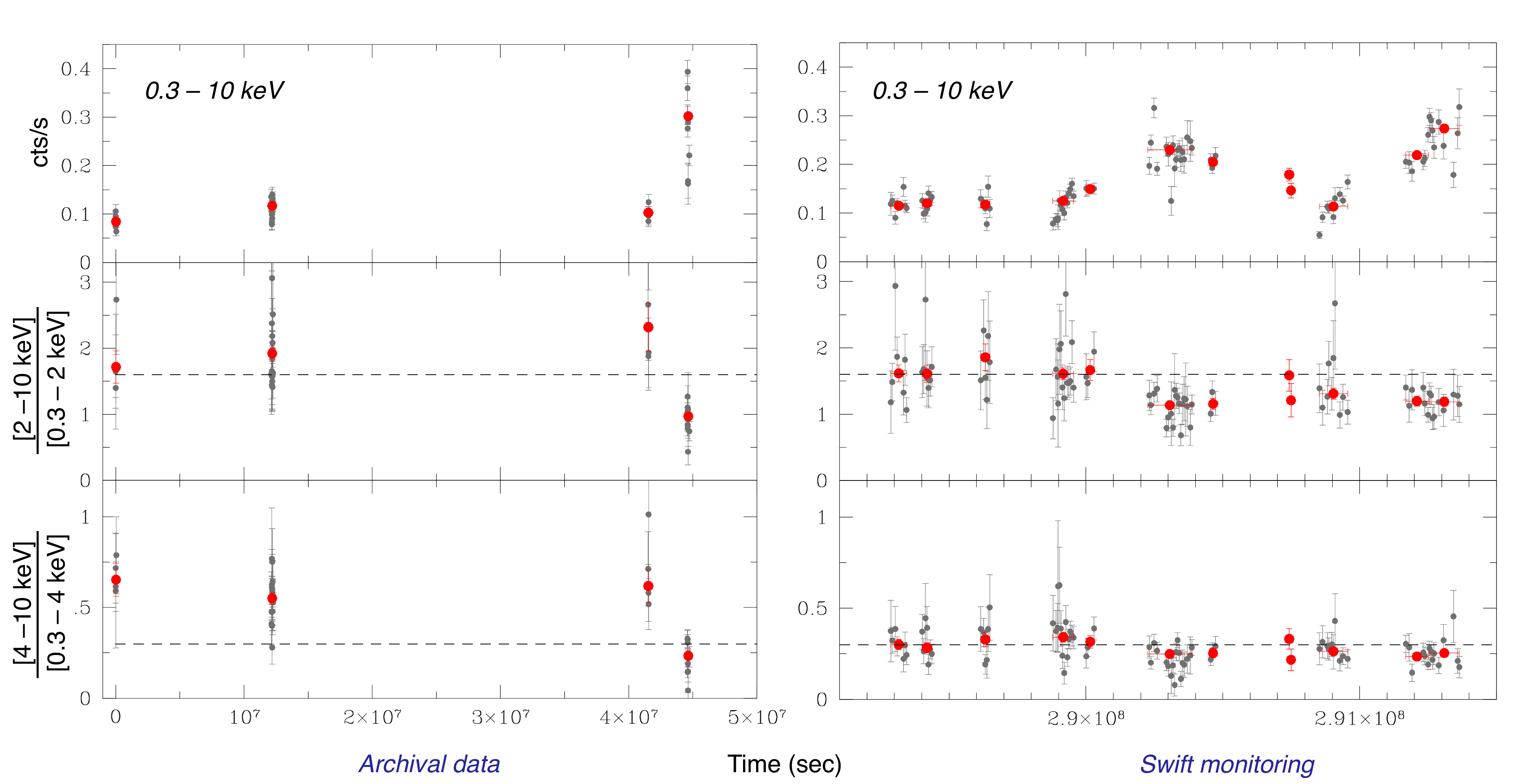}
\vspace{-0.7cm}
\caption{\footnotesize
Light curves of the $0.3-10\,$keV count rate ({\it upper panels}) and of the count rate ratios in the energy ranges $2-10\,$keV vs. $0.3-2\,$keV ({\it middle panels})
and $4-10\,$keV vs. $0.3-4\,$keV ({\it lower panels}).
We compare the trend observed in the archival data of \mrk\ ({\it left panels}) with the results from the September 2014 \swift\ monitoring ({\it right panels}).
As time zero-point, the starting time of the first archival observation is assumed.
Grey dots: one bin per snapshot; red filled circles: one bin per observation. 
To highlight the observed variation in the $H/S$, we mark with grey dashed lines the mean ratio observed in the first observation of the monitoring.
From Severgnini et al. (in prep.).
}
\vspace{-0.3cm}
\label{fig:mrk}
\end{figure}

We were awarded 12 observations with XRT (covering $\sim 3\,$weeks, for a total of $76~$ksec net exposure time), to further investigate the \nhsym\ variability and its timescales. 
In Fig.~\ref{fig:mrk} we show the temporal behaviour of \mrk, as reconstructed considering all the \swift\ datasets (archival on the left panel, new monitoring on the right panel). 
The data are binned both per observation (red filled circles) and per snapshot (grey dots).
Comparing the $0.3-10\,$keV light curves from the new monitoring and the archival data, 
we confirm, on shorter timescales, the count rate 
variability previously detected (see Fig.~\ref{fig:mrk}, upper panels).
However, the analysis of the ratio of count rates observed in hard and soft bands ($H/S$) highlights a change in the behaviour of the source: 
in the archival observations \mrk\ underwent to a dramatic spectral variation on a timescale of $1\,$month. 
This is clearly revealed by the decreasing of the $4-10\,$keV vs. $0.3-4\,$keV ratio of a factor of $3$ (see Fig.~\ref{fig:mrk}, lower panels).
A slightly less intense variation is also observed in the $2-10\,$keV vs. $0.3-2\,$keV ratio (about a factor of $2.4$).
During the new \swift\ monitoring the change in the spectral shape is less prominent, of a maximum factor of $\sim 1.5$ on timescales of $\sim 3.5-4\,$days.
No difference in the maximum amplitude of variation is observed when comparing the light curve of the $4-10\,$keV vs. $0.3-4\,$keV ratio with the light curve of the $2-10\,$keV vs. $0.3-2\,$keV ratio.

To summarise, the \swift\ monitoring allowed us to find \mrk\ in a different state than previously detected, and to follow the source during a spectral variation.
The origin of this behaviour 
is still under investigation;
a detailed spectral analysis will be presented in Severgnini et al. (in prep.).

\vspace{-0.3cm}
\acknowledgments
\vspace{-0.4cm}
The research leading to these results has received funding from the European Commission Seventh Framework Programme 
(FP7/2007-2013) under grant agreement n.~267251 ``Astronomy Fellowships in Italy'' (AstroFIt).
Support from the Italian Space Agency is acknowledged by LB (contract ASI INAF NuSTAR I/037/12/0).
This work made use of data supplied by the UK \swift\ Science Data Centre at the University of Leicester.
This research has made use of the Palermo BAT Catalogue and database operated at INAF - IASF Palermo.
We also want to thank Neil Gehrels and the \swift\ Mission Operation Center to make every effort to get our ToO request scheduled.

\vspace{-0.3cm}
\bibliographystyle{azh}

\end{document}